\documentclass[apjl]{emulateapj}
\usepackage{comment}
\usepackage{ifthen}
\usepackage{afterpage}

\newboolean{emulateapj}
\setboolean{emulateapj}{true}

\newcommand{\ctbd}[1]{}

\newcommand{\lc}{light curve}
\newcommand{\lcs}{light curves}
\newcommand{\Lc}{Light curve}

\newcommand{\oot}{out-of-transit}

\newcommand{\band}[1]{\ensuremath{#1}-band}

\newcommand{\chisq}{\ensuremath{\chi^2}}

\newcommand{\kms}{\ensuremath{\rm km\,s^{-1}}}
\newcommand{\ms}{\ensuremath{\rm m\,s^{-1}}}

\newcommand{\gcmc}{\ensuremath{\rm g\,cm^{-3}}}

\newcommand{\trillionergscmsq}{\ensuremath{10^{12}\,\rm erg\,s^{-1}\,cm^{-2}}}

\newcommand{\masyr}{\ensuremath{\rm mas\,yr^{-1}}}

\newcommand{\vsini}{\ensuremath{v \sin{i}}}
\newcommand{\feh}{\ensuremath{\rm [Fe/H]}}

\newcommand{\vmac}{\ensuremath{v_{\rm mac}}}
\newcommand{\vmic}{\ensuremath{v_{\rm mic}}}

\newcommand{\rsun}{\ensuremath{R_\sun}}
\newcommand{\msun}{\ensuremath{M_\sun}}
\newcommand{\lsun}{\ensuremath{L_\sun}}

\newcommand{\rstar}{\ensuremath{R_\star}}
\newcommand{\mstar}{\ensuremath{M_\star}}
\newcommand{\lstar}{\ensuremath{L_\star}}

\newcommand{\teffstar}{\ensuremath{T_{\rm eff\star}}}
\newcommand{\rhostar}{\ensuremath{\rho_\star}}
\newcommand{\loggstar}{\ensuremath{\log{g_{\star}}}}

\newcommand{\rearth}{\ensuremath{R_\earth}}
\newcommand{\mearth}{\ensuremath{M_\earth}}

\newcommand{\rpl}{\ensuremath{R_{p}}}
\newcommand{\mpl}{\ensuremath{M_{p}}}

\newcommand{\rhopl}{\ensuremath{\rho_{p}}}

\newcommand{\arstar}{\ensuremath{a/\rstar}}
\newcommand{\zrstar}{\ensuremath{\zeta/\rstar}}

\newcommand{\rjup}{\ensuremath{R_{\rm J}}}
\newcommand{\mjup}{\ensuremath{M_{\rm J}}}

\newcommand{\reffig}[1]{Fig.~\ref{fig:#1}}
\newcommand{\refsec}[1]{\mbox{\S\ \ref{sec:#1}}}

\newcommand{\reftab}[1]{Tab.~\ref{tab:#1}}

\newcommand{\flwof}{\mbox{FLWO 1.2\,m}}

\newcommand{\flwos}{\mbox{FLWO 1.5\,m}}

\newcommand{\hatcurfield}{145}                                         
\newcommand{\hatcurCCtwomass}{2MASS~13573347+4329367}                  
\newcommand{\hatcurCCgsc}{GSC~03033-00706}                             
\newcommand{\hatcurCCtassmv}{12.84}                                   
\newcommand{\hatcurCCtwomassJmag}{\ensuremath{10.794\pm0.023}}         
\newcommand{\hatcurCCtwomassHmag}{\ensuremath{10.236\pm0.022}}         
\newcommand{\hatcurCCtwomassKmag}{\ensuremath{10.108\pm0.016}}         
\newcommand{\hatcurCCcitJmag}{\ensuremath{10.794\pm0.024}}             
\newcommand{\hatcurCCcitHmag}{\ensuremath{10.229\pm0.023}}             
\newcommand{\hatcurCCcitKmag}{\ensuremath{10.132\pm0.017}}             
\newcommand{\hatcurLCdip}{\ensuremath{20}}                             
\newcommand{\hatcurLCrprstar}{\ensuremath{0.1406\pm0.0013}}            
\newcommand{\hatcurLCbsq}{\ensuremath{0.044_{-0.024}^{+0.035}}}        
\newcommand{\hatcurLCimp}{\ensuremath{0.211_{-0.078}^{+0.066}}}        
\newcommand{\hatcurLCzeta}{\ensuremath{23.57\pm0.12}}                  
\newcommand{\hatcurLCdur}{\ensuremath{0.0974\pm0.0006}}                
\newcommand{\hatcurLCdurhr}{\ensuremath{2.337\pm0.015}}                
\newcommand{\hatcurLCq}{\ensuremath{0.0303\pm0.0002}}                  
\newcommand{\hatcurLCingdur}{\ensuremath{0.0125\pm0.0005}}             
\newcommand{\hatcurLCP}{\ensuremath{3.2130598\pm0.0000021}}            
\newcommand{\hatcurLCPprec}{\ensuremath{3.2130598}}                    
\newcommand{\hatcurLCPshort}{\ensuremath{3.2131}}                      
\newcommand{\hatcurLCT}{\ensuremath{2454419.19556\pm0.00020}}          
\newcommand{\hatcurLCTA}{\ensuremath{2453792.64889\pm0.00044}}         
\newcommand{\hatcurLCTB}{\ensuremath{2454897.94147\pm0.00038}}         
\newcommand{\hatcurSMEiteff}{\ensuremath{4650\pm60}}                   
\newcommand{\hatcurSMEizfeh}{\ensuremath{-0.29\pm0.05}}                
\newcommand{\hatcurSMEilogg}{\ensuremath{4.75\pm0.10}}                 
\newcommand{\hatcurSMEivsin}{\ensuremath{0.5\pm0.4}}                  
\newcommand{\hatcurSMEivmac}{\ensuremath{2.26\pm0.0}}                  
\newcommand{\hatcurSMEivmic}{\ensuremath{0.85\pm0.0}}                  
\newcommand{\hatcurSMEiiteff}{\ensuremath{4591\pm60}}                  
\newcommand{\hatcurSMEiizfeh}{\ensuremath{-0.36\pm0.04}}               
\newcommand{\hatcurSMEiilogg}{\ensuremath{4.61\pm0.0}}                 
\newcommand{\hatcurSMEiivsin}{\ensuremath{1.73\pm0.5}}                 
\newcommand{\hatcurSMEiivmac}{\ensuremath{2.2\pm0.0}}                  
\newcommand{\hatcurSMEiivmic}{\ensuremath{0.85\pm0.0}}                 
\newcommand{\hatcurDSteff}{\ensuremath{4500\pm250}}                    
\newcommand{\hatcurDSlogg}{\ensuremath{4.0\pm0.2}}                     
\newcommand{\hatcurDSgamma}{\ensuremath{-40.51\pm0.21}}                
\newcommand{\hatcurDSnumspec}{\ensuremath{8}}                          
\newcommand{\hatcurLBiz}{\ensuremath{0.3432}}                          
\newcommand{\hatcurLBiiz}{\ensuremath{0.2493}}                         
\newcommand{\hatcurLBii}{\ensuremath{0.4323}}                          
\newcommand{\hatcurLBiii}{\ensuremath{0.2269}}                         
\newcommand{\hatcurLBig}{\ensuremath{0.8431}}                          
\newcommand{\hatcurLBiig}{\ensuremath{-0.0064}}                        
\newcommand{\hatcurISOm}{\ensuremath{0.73\pm0.02}}                     
\newcommand{\hatcurISOmlong}{\ensuremath{0.733\pm0.018}}               
\newcommand{\hatcurISOr}{\ensuremath{0.70_{-0.01}^{+0.02}}}            
\newcommand{\hatcurISOrlong}{\ensuremath{0.701_{-0.012}^{+0.017}}}     
\newcommand{\hatcurISOlogg}{\ensuremath{4.61\pm0.01}}                  
\newcommand{\hatcurISOlum}{\ensuremath{0.21_{-0.01}^{+0.02}}}          
\newcommand{\hatcurISOmv}{\ensuremath{6.89\pm0.11}}                    
\newcommand{\hatcurISOage}{\ensuremath{2.5\pm2.0}}                     
\newcommand{\hatcurISOMK}{\ensuremath{4.36\pm0.05}}                    
\newcommand{\hatcurISOspec}{K4}                                        
\newcommand{\hatcurRVK}{\ensuremath{35.8\pm1.9}}                       
\newcommand{\hatcurRVk}{\ensuremath{0}}                                
\newcommand{\hatcurRVh}{\ensuremath{0}}                                
\newcommand{\hatcurRVgamma}{\ensuremath{32.6\pm1.3}}                   
\newcommand{\hatcurRVjitter}{\ensuremath{4.82}}                        
\newcommand{\hatcurRVeccen}{\ensuremath{0}}                            
\newcommand{\hatcurRVkecc}{\ensuremath{0.052\pm0.025}}                 
\newcommand{\hatcurRVhecc}{\ensuremath{0.007\pm0.046}}                 
\newcommand{\hatcurPPi}{\ensuremath{89.0\pm0.4}}                       
\newcommand{\hatcurPPlogg}{\ensuremath{2.75\pm0.03}}                   
\newcommand{\hatcurPPar}{\ensuremath{11.77_{-0.21}^{+0.15}}}           
\newcommand{\hatcurPParel}{\ensuremath{0.0384\pm0.0003}}               
\newcommand{\hatcurPPrho}{\ensuremath{0.295\pm0.025}}                  
\newcommand{\hatcurPPmlong}{\ensuremath{0.211\pm0.012}}                
\newcommand{\hatcurPPmelong}{\ensuremath{67.02\pm3.71}}                
\newcommand{\hatcurPPrlong}{\ensuremath{0.959_{-0.021}^{+0.029}}}      
\newcommand{\hatcurPPrelong}{\ensuremath{10.75_{-0.24}^{+0.32}}}       
\newcommand{\hatcurPPmrcorr}{\ensuremath{0.63}}                        
\newcommand{\hatcurPPteff}{\ensuremath{963\pm16}}                      
\newcommand{\hatcurPPtheta}{\ensuremath{0.023\pm0.001}}                
\newcommand{\hatcurPPfluxavgnoexp}{\ensuremath{1.91\pm0.12}}      
\newcommand{\hatcurXdist}{\ensuremath{142.5_{-3.3}^{+4.2}}}            
\newcommand{\hatcurCCpm}{\ensuremath{137.36\pm2.23}}                   

\newcommand{\hatcur}{HAT-P-12}
\newcommand{\hatcurb}{HAT-P-12b}
\newcommand{\hatcurCCra}{\ensuremath{13^{\mathrm h}57^{\mathrm m}33.48^{\mathrm s}}}  
\newcommand{\hatcurCCdec}{\ensuremath{+43^{\mathrm d}29^{\mathrm m}36.7^{\mathrm s}}} 
\newcommand{\hatcurSMEversion}{i}                                      
\newcommand{\hatcurSMEteff}{\ifthenelse{\equal{\hatcurSMEversion}{i}}{\hatcurSMEiteff}{\hatcurSMEiiteff}}
\newcommand{\hatcurSMEzfeh}{\ifthenelse{\equal{\hatcurSMEversion}{i}}{\hatcurSMEizfeh}{\hatcurSMEiizfeh}}
\newcommand{\hatcurSMElogg}{\ifthenelse{\equal{\hatcurSMEversion}{i}}{\hatcurSMEilogg}{\hatcurSMEiilogg}}
\newcommand{\hatcurSMEvsin}{\ifthenelse{\equal{\hatcurSMEversion}{i}}{\hatcurSMEivsin}{\hatcurSMEiivsin}}
\newcommand{\hatcurSMEvmac}{\ifthenelse{\equal{\hatcurSMEversion}{i}}{\hatcurSMEivmac}{\hatcurSMEiivmac}}
\newcommand{\hatcurSMEvmic}{\ifthenelse{\equal{\hatcurSMEversion}{i}}{\hatcurSMEivmic}{\hatcurSMEiivmic}}

\shortauthors{Hartman et al.}
\shorttitle{\hatcur\lowercase{b}}

\begin{document}

\ifthenelse{\boolean{emulateapj}}{
\title{\hatcur\lowercase{b}: A Low-density sub-Saturn mass planet transiting a metal-poor K dwarf \altaffilmark{$\dagger$}}}
{\title{\hatcur\lowercase{b}: A Low-density sub-Saturn mass planet transiting a metal-poor K dwarf \altaffilmark{\dagger}}}

\author{
	J.~D.~Hartman\altaffilmark{1},
	G.~\'A.~Bakos\altaffilmark{1,2},
	G.~Torres\altaffilmark{1},
	G\'eza~Kov\'acs\altaffilmark{3},
	R.~W.~Noyes\altaffilmark{1},
	A.~P\'al\altaffilmark{1,3,4},
	D.~W.~Latham\altaffilmark{1},
	B.~Sip\H{o}cz\altaffilmark{1,4},
	D.~A.~Fischer\altaffilmark{5},
	J.~A.~Johnson\altaffilmark{6},
	G.~W.~Marcy\altaffilmark{7},
	R.~P.~Butler\altaffilmark{8},
	A.~W.~Howard\altaffilmark{7},
	G.~A.~Esquerdo\altaffilmark{1},
	D.~D.~Sasselov\altaffilmark{1},
	G\'abor~Kov\'acs\altaffilmark{1},
	R.~P.~Stefanik\altaffilmark{1},
        J.~M.~Fernandez\altaffilmark{1,9},
	J.~L\'az\'ar\altaffilmark{10},
	I.~Papp\altaffilmark{10},
	P.~S\'ari\altaffilmark{10}
}
\altaffiltext{1}{Harvard-Smithsonian Center for Astrophysics,
	Cambridge, MA, gbakos@cfa.harvard.edu}

\altaffiltext{2}{NSF Fellow}

\altaffiltext{3}{Konkoly Observatory, Budapest, Hungary}

\altaffiltext{4}{Department of Astronomy,
	E\"otv\"os Lor\'and University, Budapest, Hungary.}

\altaffiltext{5}{Department of Physics and Astronomy, San Francisco
	State University, San Francisco, CA}

\altaffiltext{6}{Institute for Astronomy, University of Hawaii,
Honolulu, HI 96822; NSF Postdoctoral Fellow}

\altaffiltext{7}{Department of Astronomy, University of California,
	Berkeley, CA}

\altaffiltext{8}{Department of Terrestrial Magnetism, Carnegie
	Institute of Washington, DC}

\altaffiltext{9}{Department of Astronomy, Pontificia Universidad
Cat\'{o}lica, Santiago, Chile}

\altaffiltext{10}{Hungarian Astronomical Association, Budapest, 
	Hungary}

\altaffiltext{$\dagger$}{%
	Based in part on observations obtained at the W.~M.~Keck
	Observatory, which is operated by the University of California and
	the California Institute of Technology. Keck time has been
	granted by NOAO (A264Hr,A146Hr) and NASA (N162Hr,N128Hr).
}

\begin{abstract}

We report on the discovery of \hatcurb{}, a transiting extrasolar
planet orbiting the moderately bright $V \approx 12.8$
\hatcurISOspec\ dwarf \hatcurCCgsc, with a period $P=\hatcurLCP$\,d,
transit epoch $T_c = \hatcurLCT$ (BJD) and transit duration
\hatcurLCdur\,d.  The host star has a mass of \hatcurISOm\,\msun,
radius of \hatcurISOr\,\rsun, effective temperature \hatcurSMEteff\,K
and metallicity $[Fe/H] = \hatcurSMEzfeh$. We find a slight
correlation between the observed spectral line bisector spans and the
radial velocity, so we consider, and rule out, various blend
configurations including a blend with a background eclipsing binary,
and hierarchical triple systems where the eclipsing body is a star or
a planet. We conclude that a model consisting of a single star with a
transiting planet best fits the observations, and show that a likely
explanation for the apparent correlation is contamination from
scattered moonlight. Based on this model, the planetary companion has
a mass of \hatcurPPmlong\,\mjup, and radius of
\hatcurPPrlong\,\rjup\ yielding a mean density of
\hatcurPPrho\,\gcmc. Comparing these observations with recent
theoretical models we find that \hatcurb{} is consistent with a $\sim
1-4.5$~Gyr, mildly irradiated, H/He dominated planet with a core mass
$M_{C} \la 10\mearth$. \hatcurb{} is thus the least massive H/He
dominated gas giant planet found to date. This record was previously
held by Saturn.
\end{abstract}


\section{Introduction}
\label{sec:introduction}

Transiting extrasolar planets (TEPs) provide unique opportunities to
study the physical properties of planetary mass objects outside of the
Solar System. By combining time-series photometric observations taken
during transit with radial velocity (RV) measurements of the star, it
is possible to precisely measure the mass and radius of the planet, if
the stellar mass and radius can be determined by other means. The bulk
density of the planet may then be compared with the predictions of
theoretical planetary structure models \citep[e.g.][]{baraffe:2008,
  fortney:2007, burrows:2007, seager:2007} to infer the structure of
the planet. Discoveries of planets that fall outside the predicted
mass-radius range \citep[e.g. inflated hot Jupiters such as
  TrES-4b;][]{mandushev:2007} lead in turn to refinements of these
models. TEPs also provide unique opportunities to study planetary
atmospheres, including their composition \citep[e.g.][]{dc:2002}, and
their thermal profiles \citep[e.g.][]{knutson:2008}. It is also
possible to measure the projection of the angle between the orbital
axis and the stellar spin axis for these planets
\citep[e.g.][]{winn:2005}, which may be used to test theories of
planetary migration \citep[][]{fabrycky:2009}.

To date more than 40 TEP discoveries have been published, with the
majority coming from dedicated photometric
surveys\footnote{http://www.exoplanet.edu/catalog-transit.php}. These
planets span a range covering more than two orders of magnitude in
mass from a Super-Earth TEP \citep[Corot-7b;][]{leger:2009} and
Super-Neptunes \citep[GJ 436b and HAT-P-11b;][]{gillon:2007,
  bakos:2009} to brown dwarf size objects \citep[Corot-3b,
  XO-3b;][]{deleuil:2008, johns-krull:2008}. Focusing on the low-mass
end, we note that the three least massive TEPs with well determined
masses \citep[GJ 436b, HAT-P-11b and HD 149026b,][]{sato:2005}, are
also the three TEPs with the smallest radii (excluding Corot-3b) and
highest inferred core mass fractions. Above this, we begin to see
planets with a wide range of radii. The planets WASP-11/HAT-P-10b
\citep{bakos:2008, west:2008}, WASP-6b \citep{gillon:2009}, HAT-P-1b
\citep{bakos:2007b}, OGLE-TR-111b \citep{pont:2004}, WASP-15b
\citep{west:2009}, and XO-2b \citep{burke:2007} all have radii larger
than or comparable to that of Jupiter, whereas HAT-P-3b
\citep{torres:2007, torres:2008} has a radius that is only slightly
larger than that of Saturn. Given the small number of TEPs known with
$M \la 0.5\mjup$, we cannot yet say what is the empirical minimum mass
of core-less, or envelope dominated, gas giant planets. To explore the
possible transition from envelope dominated to core dominated planets,
it is necessary to find more low-mass TEPs.

The Hungarian-made Automated Telescope Network
\citep[HATNet][]{bakos:2004} survey has been a major contributor to
the discovery of TEPs.  Operational since 2003, it has covered
approximately 10\% of the Northern sky, searching for TEPs around
bright stars ($8\lesssim I \lesssim 12.5$\,mag). HATNet operates six
wide field instruments: four at the Fred Lawrence Whipple Observatory
(FLWO) in Arizona, and two on the roof of the Submillimeter Array
hangar (SMA) of SAO in Hawaii. Since 2006, HATNet has announced and
published 11 TEPs. In this work we report on the 12th such
discovery. This planet is only the fourth sub-Saturn mass TEP
announced to date, but unlike the other planets, it is of low density,
and appears to be H/He dominated.

The structure of the paper is as follows. In \refsec{obs} we summarize
the observations, including the photometric detection, and follow-up
observations. In \refsec{analysis} we describe the analysis of the
data, such as the stellar parameter determination (\refsec{stelparam}),
blend modeling (\refsec{blend}), and global modeling of the data
(\refsec{globmod}). We discuss our findings in \refsec{discussion}.

\section{Observations}
\label{sec:obs}

\subsection{Photometric detection}
\label{sec:detection}

\begin{figure}[!ht]
\plotone{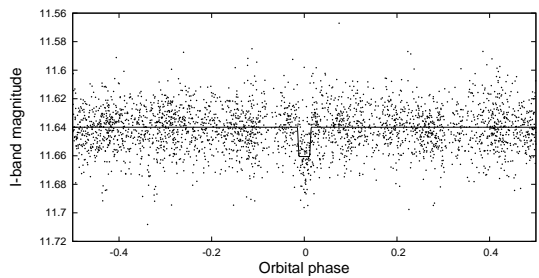}
\caption{
	The unbinned \lc{} of \hatcur{} including all 2927
        instrumental \band{I} 5.5-min cadence measurements obtained
        with the HAT-5 (Arizona) telescope of HATNet (see text for
        details), and folded with the period of $P =
        \hatcurLCPprec$\,days (which is the result of the fit
        described in \refsec{analysis}). The solid line shows a boxcar
        transit model fit to the \lc.
\label{fig:hatnet}}
\end{figure}

The transits of \hatcurb{} were detected with the HAT-5 telescope in
Arizona. The region around \hatcurCCgsc{}, a field internally labeled
as \hatcurfield, was observed on a nightly basis between January and
July of 2006, whenever weather conditions permitted. We gathered 4205
exposures of 5 minutes at a 5.5-minute cadence, of which 2927 images
were used in the final \lc\ of \hatcur{}. Each image contained
approximately 10,000 stars down to $I\sim14.0$. For the brightest
stars in the field we achieved a per-image photometric precision of
3\,mmag.

The calibration of the HATNet frames was done utilizing standard
procedures. The calibrated frames were then subjected to star
detection and astrometry, as described in \cite{pal:2006}. Aperture
photometry was performed on each image at the stellar centroids
derived from the 2MASS catalog \citep{skrutskie:2006} and the
individual astrometrical solutions.  The resulting \lcs\ were
decorrelated against trends using the External Parameter Decorrelation
technique in ``constant'' mode \citep[EPD, see][]{bakos:2009} and the
Trend Filtering Algorithm \citep[TFA, see][]{kovacs:2005}. The \lcs{}
were searched for periodic box-like signals using the Box Least
Squares method \citep[BLS, see][]{kovacs:2002}. We detected a
significant signal in the \lc{} of \hatcurCCgsc{} (also known as
\hatcurCCtwomass{}; $\alpha = \hatcurCCra$, $\delta = \hatcurCCdec$;
J2000; $V=\hatcurCCtassmv \pm 0.09$, \citealp{droege:2006}), with a depth of
$\sim\hatcurLCdip$\,mmag, and a period of $P=\hatcurLCPshort$\,days.
The dip had a relative duration (first to last contact) of
$q\approx\hatcurLCq$, corresponding to a total duration of
$Pq\approx\hatcurLCdurhr$~hours (see \reffig{hatnet}).

\subsection{Reconnaissance Spectroscopy}
\label{sec:recspec}

All HATNet candidates are subjected to thorough investigation before
using more precious time on large telescopes. One of the important
tools for establishing whether the transit-like feature in the light
curve of a candidate is due to astrophysical phenomena other than a
planet transiting a star is the CfA Digital Speedometer
\citep[DS;][]{latham:1992}, mounted on the \flwos\ telescope. This
yields high-resolution (${\rm R}=35,000$) spectra with low signal-to-noise
ratio sufficient to derive radial velocities with moderate precision
(roughly 0.5-1\,\kms), and to determine the effective temperature and
surface gravity of the host star. With this facility we are able to
reject many types of false positives, such as F dwarfs orbited by M
dwarfs, grazing eclipsing binaries, triple and quadruple star systems,
or giant stars where the transit signal cannot be due to a planet.

We obtained \hatcurDSnumspec{} observations of \hatcur{} with the
DS. The RV measurements of \hatcur{} showed an rms residual of
0.43\,\kms, consistent with no detectable RV variation. The spectra
were single-lined, showing no detectable evidence for more than one
star in the system.  Atmospheric parameters for the star, including
the effective temperature $\teffstar = \hatcurDSteff\,K$, surface
gravity $\loggstar = \hatcurDSlogg$ (log cgs), and projected
rotational velocity $\vsini$ consistent with zero with an asymmetric
error of about 1\kms, were derived as described by \cite{torres:2002}.
The effective temperature and surface gravity correspond to a mid-K
dwarf. The mean heliocentric radial velocity of \hatcur\ is
\hatcurDSgamma\,\kms.

\subsection{High resolution, high S/N spectroscopy}
\label{sec:hispec}

Given the significant transit detection by HATNet, and the positive DS
results that exclude obvious false positives, we proceeded with the
follow-up of this candidate by obtaining high-resolution and high S/N
spectra to characterize the radial velocity variations and to
determine the stellar parameters with higher precision. Using the
HIRES instrument \citep{vogt:1994} on the Keck~I telescope located on
Mauna Kea, Hawaii, we obtained 22 exposures with an iodine cell, plus
one iodine-free template. The observations were made on 16 nights
during a number of observing runs between 2007 March 27 and 2008
September 17.

\begin{figure} [ht]
\plotone{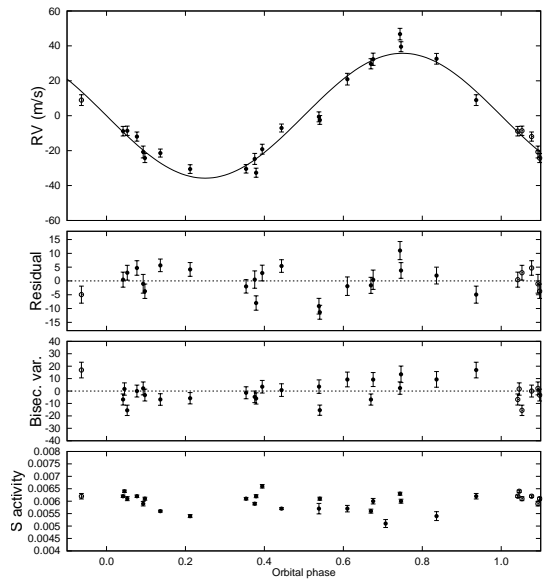}
\caption{
	Top: Radial-velocity measurements from Keck for \hatcur{},
        along with an orbital fit, shown as a function of orbital
        phase, using our best fit period (see
        \refsec{analysis}). Zero-phase is defined by the transit
        midpoint. The center-of-mass velocity has been
        subtracted. Note that the error bars show the formal errors
        given by the spectrum reduction pipeline and do not include
        our estimate of the stellar jitter.
	Second panel: Phased residuals after subtracting the orbital fit
	(also see \refsec{analysis}). The rms variation of the residuals is
	about $\hatcurRVjitter$\,\ms.
	Third panel: Bisector spans (BS) including the template
        spectrum (\refsec{blend}).  The mean value has been subtracted.
	Bottom: Relative S values for the Keck spectra. 
Note the different vertical scales of the panels.
\label{fig:rvbis}}
\end{figure}

The width of the spectrometer slit used on HIRES was $0\farcs86$,
resulting in a resolving power of $\lambda/\Delta\lambda \approx
55,\!000$, with a wavelength coverage of $\sim3800-8000$\,\AA\@. The
iodine gas absorption cell was used to superimpose a dense forest of
$\mathrm{I}_2$ lines on the stellar spectrum and establish an accurate
wavelength fiducial \citep[see][]{marcy:1992}. Relative RVs in the
Solar System barycentric frame were derived as described by
\cite{butler:1996}, incorporating full modeling of the spatial and
temporal variations of the instrumental profile. The final RV data and
their errors are listed in \reftab{rvs}. The folded data, with our
best fit (see \refsec{analysis}) superimposed, are plotted in
\reffig{rvbis}. 

In \reffig{rvbis} we also plot the relative S index. This index is
computed following the prescription given by \citet{vaughan:1978}
after matching each spectrum to a reference spectrum using a
transformation that includes a wavelength shift and a flux scaling
that is a polynomial as a function of wavelength. The transformation
is determined on regions of the spectra that are not used in computing
the S index. Note that the relative S index has not been calibrated to
the scale of \citet{vaughan:1978}. The relative S index does not show
any significant variation correlated with the orbital phase; such a
correlation might have indicated that the RV variations are due to
stellar activity.

\ifthenelse{\boolean{emulateapj}}{\begin{deluxetable*}{lrrrrrr}}{\begin{deluxetable}{lrrrrrr}}
\tablewidth{0pc}
\tablecaption{
	Relative radial velocity, bisector and activity index
	measurements of \hatcur{}.
	\label{tab:rvs}
}
\tablehead{
	\colhead{BJD} & 
	\colhead{RV\tablenotemark{a}} & 
	\colhead{\ensuremath{\sigma_{\rm RV}}\tablenotemark{b}} & 
	\colhead{BS} & 
	\colhead{\ensuremath{\sigma_{\rm BS}}} & 
	\colhead{S\tablenotemark{c}} & 
	\colhead{\ensuremath{\sigma_{\rm S}}}\\
	\colhead{\hbox{~~~~(2,454,000$+$)~~~~}} & 
	\colhead{(\ms)} & 
	\colhead{(\ms)} &
	\colhead{(\ms)} &
    \colhead{(\ms)} &
	\colhead{} &
	\colhead{}
}
\startdata
$ 187.03115 $ \dotfill & $    46.75 $ & $     3.29 $ & $     2.38 $ & $     4.99 $ & $    0.0063 $ & $   0.00006 $\\
$ 187.98977 $ \dotfill & $    -8.83 $ & $     2.74 $ & $    -6.80 $ & $     4.45 $ & $    0.0062 $ & $   0.00005 $\\
$ 188.00121\tablenotemark{d} $ \dotfill & \nodata      & \nodata      & $     1.64 $ & $     4.97 $ & $    0.0064 $ & $   0.00005 $\\
$ 188.10261 $ \dotfill & $   -11.93 $ & $     2.63 $ & $    -0.01 $ & $     4.82 $ & $    0.0062 $ & $   0.00005 $\\
$ 188.15398 $ \dotfill & $   -20.76 $ & $     3.39 $ & $     2.13 $ & $     5.21 $ & $    0.0059 $ & $   0.00009 $\\
$ 188.99152 $ \dotfill & $   -30.42 $ & $     2.43 $ & $    -1.37 $ & $     4.82 $ & $    0.0061 $ & $   0.00005 $\\
$ 189.07296 $ \dotfill & $   -32.64 $ & $     2.61 $ & $    -5.94 $ & $     4.52 $ & $    0.0062 $ & $   0.00006 $\\
$ 189.12279 $ \dotfill & $   -19.16 $ & $     2.86 $ & $     3.48 $ & $     5.09 $ & $    0.0066 $ & $   0.00007 $\\
$ 216.94159 $ \dotfill & $    -8.61 $ & $     2.68 $ & $   -15.51 $ & $     4.19 $ & $    0.0061 $ & $   0.00008 $\\
$ 247.86057 $ \dotfill & $    32.36 $ & $     3.50 $ & $     9.26 $ & $     5.66 $ & $    0.0060 $ & $   0.00011 $\\
$ 250.86401 $ \dotfill & $    20.94 $ & $     3.35 $ & $     9.31 $ & $     5.95 $ & $    0.0057 $ & $   0.00013 $\\
$ 251.91230 $ \dotfill & $     8.95 $ & $     3.09 $ & $    16.91 $ & $     6.31 $ & $    0.0062 $ & $   0.00011 $\\
$ 548.03028 $ \dotfill & $   -24.24 $ & $     2.61 $ & $    -3.24 $ & $     4.77 $ & $    0.0061 $ & $   0.00006 $\\
$ 548.92425 $ \dotfill & $   -24.73 $ & $     3.15 $ & $    -4.65 $ & $     4.66 $ & $    0.0059 $ & $   0.00005 $\\
$ 602.77839 $ \dotfill & $   -21.39 $ & $     2.29 $ & $    -6.79 $ & $     4.64 $ & $    0.0056 $ & $   0.00004 $\\
$ 603.02042 $ \dotfill & $   -30.57 $ & $     2.49 $ & $    -5.78 $ & $     4.54 $ & $    0.0054 $ & $   0.00006 $\\
$ 603.76466 $ \dotfill & $    -7.05 $ & $     2.28 $ & $     0.76 $ & $     5.08 $ & $    0.0057 $ & $   0.00004 $\\
$ 604.06935 $ \dotfill & $    -0.63 $ & $     2.84 $ & $     3.57 $ & $     5.40 $ & $    0.0057 $ & $   0.00021 $\\
$ 633.94384 $ \dotfill & $    32.62 $ & $     3.05 $ & $     9.37 $ & $     6.30 $ & $    0.0054 $ & $   0.00018 $\\
$ 636.86807 $ \dotfill & $    39.55 $ & $     2.84 $ & $    13.46 $ & $     6.64 $ & $    0.0060 $ & $   0.00008 $\\
$ 639.83599 $ \dotfill & $    29.74 $ & $     2.92 $ & $    -6.83 $ & $     4.50 $ & $    0.0056 $ & $   0.00008 $\\
$ 674.76312 $ \dotfill & $    -2.42 $ & $     2.52 $ & $   -15.39 $ & $     4.13 $ & $    0.0061 $ & $   0.00006 $\\
\enddata
\tablenotetext{a}{
	The fitted zero-point that is on an arbitrary scale has {\em not} been
	subtracted from the velocities.
}
\tablenotetext{b}{
	The values for $\sigma_{\rm RV}$ are the formal uncertainties from the spectrum reduction pipeline and do not include our estimate of the stellar jitter.
}
\tablenotetext{c}{
	This is a relative S index that is sensitive to variations in
        S, it has not been calibrated to the scale of
        \citet{vaughan:1978}.
}
\tablenotetext{d}{
        This is the iodine-free template exposure for which we do not measure 
        the RV but do measure the BS and S index.
}
\ifthenelse{\boolean{emulateapj}}{\end{deluxetable*}}{\end{deluxetable}}
\ifthenelse{\boolean{emulateapj}}{}{\clearpage}

\subsection{Photometric follow-up observations}
\label{sec:phot}

\begin{figure}[!ht]
\plotone{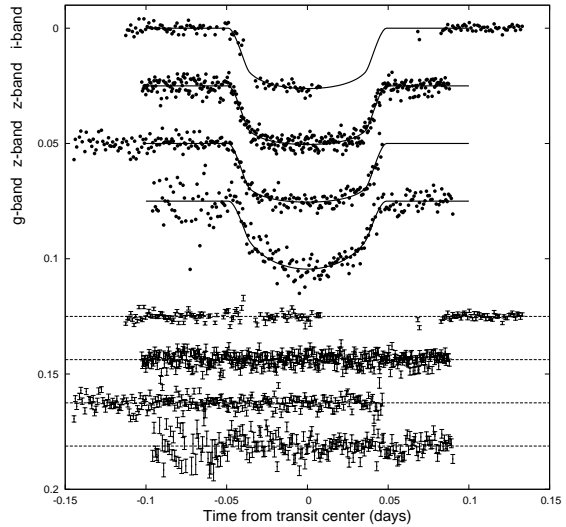}
\caption{
	Unbinned instrumental $i$-, $z$- and \band{g} transit \lcs{},
        acquired with KeplerCam at the \flwof{} telescope on the
        nights of 2007 March 27, 2007 April 25, 2009 February 5 and
        2009 March 6 from top to bottom.  Superimposed are the
        best-fit transit model \lcs.  In the bottom of the figure we
        show the residuals from the fit.  Error bars represent the
        photon and background shot-noise, plus the readout noise.
\label{fig:lc}}
\end{figure}

To confirm the transit signal and obtain high-precision light curves
for modeling the system we conducted photometric follow-up
observations with the KeplerCam CCD on the \flwof{} telescope. We
observed four transit events of \hatcurb{} on the nights of 2007 March
27, 2007 April 25, 2009 February 5 and 2009 March 6
(Fig.~\ref{fig:lc}). On 2007 March 27, 151 frames were acquired with a
cadence of 90 seconds (75 seconds of exposure time) in the Sloan
\band{i}; the observations were interrupted at mid-transit due to
clouds. On 25 April 2007, 372 frames were acquired with a cadence of
45 seconds (30 seconds of exposure time) in the Sloan \band{z}. On 5
February 2009, 218 frames were obtained with a cadence of 70 seconds
(30 seconds exposure time) in \band{z}. Finally, on 6 March 2009, 213
frames were acquired with a cadence of 70 seconds (60 seconds exposure
time) in the Sloan \band{g}. This follow-up \band{g} light curve was
obtained to further constrain possible blend scenarios
(\refsec{blend}).

The reduction of the images was performed as follows. After bias and
flat-field calibration, we derived an initial second-order
astrometric transformation between the $\sim60$ brightest stars and
the 2MASS catalog, as described in \citet{pal:2006}, yielding a
residual of $\sim0.3-0.4$ pixels. The primary reason for precise
astrometry is to minimize the photometric errors that would originate
from the centroid errors for the individual stars on each
frame. Aperture photometry was then performed on the resulting fixed
positions, using a series of apertures. The instrumental magnitude
transformation was done in two steps: first, all magnitude values were
transformed to the photometric reference frame (selected to be the
sharpest image), using the individual Poisson noise error estimates
as weights. In the second step, the magnitude fit was repeated using
the mean individual \lc{} magnitudes as reference and the rms of these
\lcs{} as weights. In both of the magnitude transformations, we
excluded from the fit the target star itself and the $3$$\sigma$
outliers. We performed EPD and TFA against trends simultaneously with
the light curve modeling (for more details, see \refsec{analysis} and
\citealp{bakos:2009}). From the series of apertures, for each night,
we chose the one yielding the smallest fit rms for the \lc. This
aperture conveniently fell in the middle of the aperture series. The
final \lcs{} are shown in the upper plots of \reffig{lc}, with our best
fit transit \lc{} models superimposed (see also \refsec{analysis}).

\begin{deluxetable}{lrrrr}
\tablewidth{0pc}
\tablecaption{Photometry follow-up of \hatcur\label{tab:phfu}}
\tablehead{
	\colhead{BJD} & 
	\colhead{Mag\tablenotemark{a}} & 
	\colhead{\ensuremath{\sigma_{\rm Mag}}} &
	\colhead{Mag(orig)\tablenotemark{b}} & 
	\colhead{Filter} \\
	\colhead{\hbox{~~~~(2,454,000$+$)~~~~}} & 
	\colhead{} & 
	\colhead{} &
	\colhead{} & 
	\colhead{}
}
\startdata
$ 187.74274 $ & $   0.00314 $ & $   0.00073 $ & $  10.51770 $ &  i\\
$ 187.74377 $ & $   0.00073 $ & $   0.00073 $ & $  10.51630 $ &  i\\
$ 187.74478 $ & $   0.00086 $ & $   0.00074 $ & $  10.51700 $ &  i\\
$ 187.74583 $ & $  -0.00138 $ & $   0.00072 $ & $  10.51430 $ &  i\\
$ 187.74683 $ & $   0.00142 $ & $   0.00073 $ & $  10.51610 $ &  i\\
$ 187.74786 $ & $   0.00369 $ & $   0.00074 $ & $  10.51970 $ &  i\\
$ 187.74888 $ & $  -0.00010 $ & $   0.00073 $ & $  10.51570 $ &  i\\
$ 187.74991 $ & $   0.00207 $ & $   0.00073 $ & $  10.51720 $ &  i\\
$ 187.75095 $ & $  -0.00404 $ & $   0.00072 $ & $  10.51180 $ &  i\\
$ 187.75196 $ & $   0.00101 $ & $   0.00072 $ & $  10.51520 $ &  i\\
\enddata
\tablenotetext{a}{
	The out-of-transit level has been subtracted. These magnitudes have
	been subjected to EPD and TFA procedure, carried out simultaneously
	with the transit fit.
}
\tablenotetext{b}{
        The raw magnitude values without application of the EPD and
        TFA procedures.
}
\tablecomments{
	This table is presented in its entirety in the electronic edition
	of the Astrophysical Journal. A portion is shown here for guidance
	regarding its form and content.
}
\end{deluxetable}

\section{Analysis}
\label{sec:analysis}

\subsection{Properties of the parent star}
\label{sec:stelparam}

We derived the initial stellar atmospheric parameters by using the
template spectrum obtained with the Keck/HIRES instrument. We used the
SME package of \cite{valenti:1996} along with the atomic-line database
of \cite{valenti:2005}, which yielded the following {\em initial}
values and uncertainties (which we have conservatively increased to
include our estimates of the systematic errors):
effective temperature $\teffstar=\hatcurSMEiteff$\,K, 
stellar surface gravity $\loggstar=\hatcurSMEilogg$\,(cgs),
metallicity $\feh=\hatcurSMEizfeh$\,dex, and 
projected rotational velocity $\vsini=\hatcurSMEivsin\,\kms$.


At this stage we could use the effective temperature and the surface
gravity as a luminosity indicator, and determine the stellar
parameters based on these two constraints using a set of
isochrones. However, the effect of \loggstar\ on the spectral line
shapes is typically subtle and as a result it is generally a rather
poor luminosity indicator in practice. For planetary transits, the
$\arstar$ normalized semi-major axis and related $\rhostar$ mean
stellar density typically impose a stronger constraint on possible
stellar models \citep{sozzetti:2007}. The validity of our assumption,
namely that the adequate physical model describing our data is a
planetary transit (as opposed to a blend), is shown later in
\refsec{blend}.

Using the values of \teffstar, \feh, and \loggstar\ from the SME
analysis, and corresponding quadratic limb darkening coefficients
($a_{z}$, $b_{z}$, etc.) from \cite{claret:2004}, we performed a
global modeling of the data (\refsec{globmod}), yielding a full
Monte-Carlo distribution of \arstar. This was complemented by a
Monte-Carlo distribution of \teffstar\ and \feh, obtained by assuming
Gaussian uncertainties based on the 1-$\sigma$ error bars of the
initial SME analysis.

\begin{figure}[!ht]
\plotone{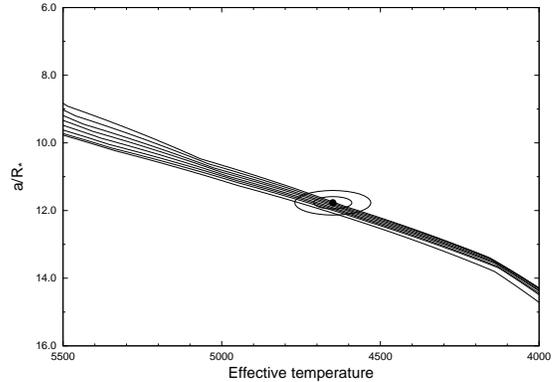}
\caption{
	Stellar isochrones from \cite{baraffe:1997,baraffe:1998} for metallicity
        \feh=$-0.29$ and ages 0.2, 0.5, 1.0, 2.0, 3.0, 4.0,
        5.0 and 6.0\,Gyr.  The final choice of $\teffstar$ and
        \arstar\ are marked and encircled by the 1-$\sigma$ and
        2-$\sigma$ confidence ellipsoids.
\label{fig:iso}}
\end{figure}

For each combination within the large ($\sim10^4$) set of \arstar,
\teffstar, and \feh\ values, we searched the stellar isochrones of the
\citet{baraffe:1997,baraffe:1998} models for the best fit stellar
model parameters (such as \mstar, \rstar, \loggstar, etc).  We
interpolated these isochrones to the SME-based stellar metallicity of
$\feh = \hatcurSMEzfeh$.
The majority of the parameter combinations in the Monte Carlo search
did not match any isochrone. In such cases ($\sim60$\% of all trials)
we skipped to the next randomly drawn parameter set. At the end we
derived the mean values and uncertainties of the physical parameters
based on their {\em a posteriori} distribution. We note that the
spread of the input stellar parameters (based on the Gaussian
uncertainties) was large compared to what the isochrones cover as a
function of age, due to the very slow evolution of K dwarfs (see
\reffig{iso}). This is partly the reason for the 40\% match ratio. We
also note that the match ratio is very sensitive to changing
fundamental parameters of the isochrones, such as the mixing length or
the metallicity.


	We then repeated the SME analysis by fixing \loggstar\ to the
        refined value of $\loggstar=\hatcurISOlogg$ based on the
        isochrone search, and only adjusting \teffstar, \feh\ and
        \vsini.  This second iteration yielded
        $\teffstar=\hatcurSMEiiteff$\,K, $\feh = \hatcurSMEiizfeh$ and
        $\vsini = \hatcurSMEiivsin$\,\kms. Curiously, the new
        \teffstar\ and \feh\ values from this second iteration provide
        a somewhat inferior match with the
        \citet{baraffe:1997,baraffe:1998} isochrones, as compared to
        the initial match. Possible reasons for this include i)
        systematic errors in the SME analysis due to the relatively
        low SNR of our Keck spectra, ii) increasing uncertainty in the
        SME analysis due to the late stellar type of the host star
        (note that \hatcur{} has a temperature that is below the
        4700~K cut-off for stars included in the analysis of
        \citealp{valenti:2005}), iii) general uncertainty in the
        isochrones for mid-K dwarfs \citep[there is a well-known
          discrepancy between the observed and predicted mass-radius
          relation for K and M dwarf stars in double-lined eclipsing
          binaries such that the observed radii are larger than the
          predicted radii, though there is some evidence that this
          discrepancy only holds for rapidly rotating, active stars,
          e.g.][]{torres:2002b,ribas:2006,lopez-morales:2007,chabrier:2007}.
        Thus we accepted the initial values of \teffstar, \feh\ and
        \vsini\ as the final atmospheric parameters for this star,
        along with the isochrone based stellar parameters, yielding
        \mstar=\hatcurISOmlong\,\msun,
        \rstar=\hatcurISOrlong\,\rsun\ and
        \lstar=\hatcurISOlum\,\lsun. Along with other stellar
        parameters, these are summarized in \reftab{stellar}.

The stellar evolutionary isochrones from
\cite{baraffe:1997,baraffe:1998} for metallicity
\feh=$-0.29$ are plotted in \reffig{iso}, with the final
choice of effective temperature $\teffstar$ and \arstar\ marked, and
encircled by the 1-$\sigma$ and 2-$\sigma$ confidence ellipsoids.


The stellar evolution modeling also yields the absolute magnitudes and
colors in various photometric passbands. We used the apparent
magnitudes from the 2MASS catalogue \citep{skrutskie:2006} to
determine the distance of the system. The magnitudes reported in the
2MASS catalogue have to be converted to the CIT system
\citep{elias:1982,elias:1983}, in which the stellar evolution models
specify the colors.
The reported magnitudes for this star are
$J_{\rm 2MASS}=\hatcurCCtwomassJmag$, 
$H_{\rm 2MASS}=\hatcurCCtwomassHmag$ and 
$K_{\rm 2MASS}=\hatcurCCtwomassKmag$;
which are equivalent to
$J_{\rm CIT}=\hatcurCCcitJmag$, 
$H_{\rm CIT}=\hatcurCCcitHmag$ and 
$K_{\rm CIT}=\hatcurCCcitKmag$;
in the CIT photometric system \citep[see][]{carpenter:2001}. Thus, the
converted 2MASS magnitudes yield a color of $(J-K)=0.66\pm0.03$ that
is within 1-$\sigma$ of the expected, isochrone-based $(J-K)_{\rm
  Baraffe}=0.60\pm0.09$. We thus relied on the 2MASS $K$ apparent
magnitude and the $M_{\rm K}=\hatcurISOMK$ absolute magnitude derived
from the above-mentioned modelling to determine the distance:
\hatcurXdist\,pc. The \band{K} was chosen because it is the longest
wavelength band-pass with the smallest expected discrepancies due to
molecular lines in the spectrum of this \hatcurISOspec\ dwarf.


\newcommand{\hatcurisoshort}{Baraffe}
\newcommand{\hatcurlumind}{\arstar}
\begin{deluxetable}{lcl}
\tablewidth{0pc}
\tablecaption{
	Stellar parameters for \hatcur{}
	\label{tab:stellar}
}
\tablehead{
	\colhead{Parameter}	&
	\colhead{Value} &
	\colhead{Source}
}
\startdata
$\teffstar$ (K)\dotfill         &  \hatcurSMEteff   & SME\tablenotemark{a}\\
$\feh$\dotfill                  &  \hatcurSMEzfeh   & SME                 \\
$\vsini$ (\kms)\dotfill         &  \hatcurSMEvsin   & SME                 \\
$\vmac$ (\kms)\dotfill          &  \hatcurSMEvmac   & SME                 \\
$\vmic$ (\kms)\dotfill          &  \hatcurSMEvmic   & SME                 \\
$\gamma_{\rm RV}$ (\kms)\dotfill         &  \hatcurDSgamma       & DS      \\
$a_{z}$\dotfill			&  \hatcurLBiz		& SME+Claret\tablenotemark{b}\\
$b_{z}$\dotfill			&  \hatcurLBiiz		& SME+Claret          \\
$a_{i}$\dotfill			&  \hatcurLBii		& SME+Claret          \\
$b_{i}$\dotfill			&  \hatcurLBiii		& SME+Claret          \\
$a_{g}$\dotfill			&  \hatcurLBig		& SME+Claret          \\
$b_{g}$\dotfill			&  \hatcurLBiig		& SME+Claret          \\
$\mstar$ ($\msun$)\dotfill      &  \hatcurISOmlong       & \hatcurisoshort+\hatcurlumind+SME\tablenotemark{c}\\
$\rstar$ ($\rsun$)\dotfill      &  \hatcurISOrlong       & \hatcurisoshort+\hatcurlumind+SME         \\
$\loggstar$ (cgs)\dotfill       &  \hatcurISOlogg    & \hatcurisoshort+\hatcurlumind+SME         \\
$\lstar$ ($\lsun$)\dotfill      &  \hatcurISOlum     & \hatcurisoshort+\hatcurlumind+SME         \\
$V$ (mag)\dotfill				&  \hatcurCCtassmv   & TASS                                      \\
$M_V$ (mag)\dotfill             &  \hatcurISOmv      & \hatcurisoshort+\hatcurlumind+SME         \\
$K$ (mag,CIT)                   &  \hatcurCCcitKmag  & 2MASS+Carpenter\tablenotemark{d}          \\
$M_K$ (mag,CIT)\dotfill         &  \hatcurISOMK      & \hatcurisoshort+\hatcurlumind+SME         \\
Age (Gyr)\dotfill               &  \hatcurISOage     & \hatcurisoshort+\hatcurlumind+SME         \\
Distance (pc)\dotfill           &  \hatcurXdist      & \hatcurisoshort+\hatcurlumind+SME\\
\enddata \tablenotetext{a}{SME = ``Spectroscopy Made Easy'' package
  for analysis of high-resolution spectra \cite{valenti:1996}. These
  parameters depend primarily on SME, with a small dependence on the
  iterative analysis incorporating the isochrone search and global
  modeling of the data, as described in the text.}
\tablenotetext{b}{SME+Claret = Based on the SME analysis and tables of
  quadratic limb darkening coefficients from \citet{claret:2004}.}
\tablenotetext{c}{\hatcurisoshort+\hatcurlumind+SME =
  \citet{baraffe:1997,baraffe:1998} isochrones, \arstar\ relative
  semi-major axis, and SME results.}  \tablenotetext{d}{Based on the
  relations from \citet{carpenter:2001}.}
\end{deluxetable}

\subsection{Excluding blend scenarios}
\label{sec:blend}

\subsubsection{Spectral line-bisector analysis}
\label{sec:bisec}

Following \cite{torres:2007}, we explored the possibility that the
measured radial velocities are not real, but are instead caused by
distortions in the spectral line profiles due to contamination from a
nearby unresolved eclipsing binary. A bisector analysis based on the
Keck spectra was done as described in \S 5 of \cite{bakos:2007a}.

Fig.~\ref{fig:rvbis} shows the bisector spans (BS) phased with the
orbital period of the planet. In calculating the BS we use the
convention
\begin{equation}
{\rm BS} = v_{\rm low} - v_{\rm high}
\end{equation}
where $v_{\rm low}$ is the velocity of the bisector of the
cross-correlation profile at a low cross-correlation value, and
$v_{\rm high}$ is the velocity at a high cross-correlation
value. While the BS do not show significant variations with an
amplitude that is comparable to or greater than the RV variations,
there does appear to be a correlation between the BS and RV
measurements. Applying a Spearman rank-order correlation test, we find
that the two variables are correlated with 98\% confidence (i.e. there
is a $2\%$ false alarm probability). Since such a correlation might
indicate a blend scenario, we consider below, and rule out, the
possibility that the system is a blend between a bright foreground K
star and a background eclipsing binary (\refsec{bgeb}) or a
hierarchical triple system (\refsec{hitrip}). The scenarios that we
consider are summarized in Table~\ref{tab:blendtypes}. We then
consider the possibility that the correlation is not astrophysical,
but rather results from variations in the sky contamination of the
spectra (\refsec{scf}); we conclude that this is the most likely
explanation for the correlation.


\subsubsection{Contamination from a background eclipsing binary}
\label{sec:bgeb}

\begin{figure}[!ht]
\plotone{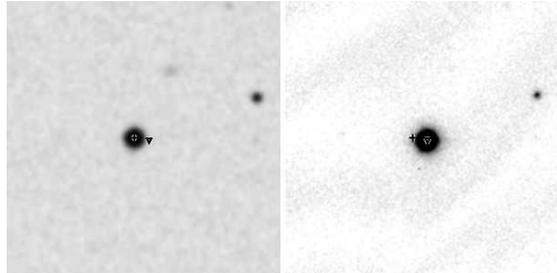}
\caption{
	Images of a $2\farcm5 \times 2\farcm5$ field containing
        \hatcur{} from the POSS-I Red survey (Left), and from our
        \flwof\ \band{z} follow-up observations (Right, see
        section~\ref{sec:phot}). The dates of the exposures are 1955
        April 13 and 2009 February 5 respectively. The cross marks the
        position of \hatcur{} in 1955 and the triangle marks the
        position in 2009. Between these two dates \hatcur{} moved
        $\sim 7\farcs5$ to the West. From the POSS-I image we can rule
        out the presence of stars brighter than $R \sim 19$ at the
        current position of \hatcur{}.
\label{fig:poss}}
\end{figure}

The high proper motion of \hatcur{} allows us to rule out one possible
scenario that could potentially fit the available observations, namely
a background eclipsing binary that is aligned, by chance, with the
foreground \hatcurISOspec\ dwarf \hatcur{} (we refer to this as the
H,b(S-s) model, where H stands for the foreground star \hatcur{}, and
the b(S-s) is a background eclipsing binary star system; here 'b'
refers to the fact that the eclipsing binary is in the background
rather than being associated with the star H). To reproduce the
observed $\sim 2.5\%$ deep transit the background object cannot be
more than 4 mag fainter than \hatcur{} (objects fainter than this
would contribute less than 2.5\% of the total combined light and so
could not cause the transit even if they were to be completely
eclipsed by an object that emits no light). Because \hatcur{} has a
high proper motion \citep[\hatcurCCpm\,\masyr;][]{zacharias.2004} it
is possible to use the Palomar Observatory Sky Survey plates from 1955
(POSS-I, red and blue plates) to view the sky at the current position
of \hatcur{} \citep[this same technique was used for
  HAT-P-11,][]{bakos:2009}. Between 1955 and the follow-up
observations in 2009, \hatcur{} has moved $\sim
7\farcs5$. Fig.~\ref{fig:poss} shows an image stamp from the POSS-I
plate compared with a recent observation from the \flwof. We can rule
out a background object down to $\sim 19$~mag within $\sim 3\arcsec$
of the current position of \hatcur{}. Any background object must be
$\ga 6$~mag fainter than \hatcur{} and thus could not be responsible
for the observed transit.

\subsubsection{Detailed modeling of a hierarchical triple}
\label{sec:hitrip}

Following \cite{bakos:2009} we consider the possibility that \hatcur{}
is a hierarchical triple system, consisting of two eclipsing bodies
that are diluted by a third star. In the following we
refer to the bright star, with properties determined from the SME
analysis, as \hatcur{}. We consider three scenarios. In the first
scenario we assume that the bright star \hatcur{} is uneclipsed, and
that the two eclipsing components are stars with parameters
constrained by common origin to fall on the same age/metallicity
isochrone as \hatcur{} (we refer to this model as the H,S-s model,
where H denotes the bright star \hatcur{} and S-s denotes a physically
associated eclipsing binary consisting of a brighter star S and a
fainter star s). In the second scenario we assume that \hatcur{} is
uneclipsed, that one of the eclipsing components is a fainter star and
that the other component is a planet with negligible mass and
luminosity compared to the star (the H,s-p model, where s stands for
the fainter star, and p is the planet). In the third scenario we
assume that \hatcur{} is a star that is transited by a planet and that
there is a fainter star diluting the observed transit (the H-p,s
model). These models will be compared to the fiducial model of a
single star orbited by a planet (the H-p model). 

\ifthenelse{\boolean{emulateapj}}{\begin{deluxetable*}{lll}}{\begin{deluxetable}{lll}}
\ifthenelse{\boolean{emulateapj}}{}{\rotate}
\tablewidth{0pc}
\tablecaption{Blend Configurations\label{tab:blendtypes}}
\tablehead{
	\colhead{Abbreviation} &
	\colhead{Description} &
        \colhead{Excluded by}
}
\startdata
H,b(S-s) & Eclipsing binary star diluted by unresolved, unrelated star & Proper Motion (\refsec{bgeb}) \\
H,S-s & Hierarchical triple star system, two components are eclipsing & Light curve fit (\refsec{hitrip}) \\
H,s-p & Binary star system, fainter star has a transiting planet & Light curve fit (\refsec{hitrip}) \\
H-p,s & Binary star system, brighter star has a transiting planet & Light curve fit (\refsec{hitrip}) \\
H-p  & Single star with a transiting planet & Not excluded \\
\enddata
\ifthenelse{\boolean{emulateapj}}{\end{deluxetable*}}{\end{deluxetable}}

For the H,S-s and H,s-p scenarios we fit the follow-up $z$-, $i$-, and
\band{g} \lcs\ together with the HATNet \band{I} \lc\ following the
procedure described by \cite{bakos:2009}. We include the HATNet
\lc\ to constrain the possibility of a secondary eclipse; to exclude
points that do not contribute to the fit, we only include points that
are within 1 transit duration of the start of transit ingress or end
of transit egress, or within 1 transit duration of the start of
secondary ingress or end of secondary egress. We use the TFA HATNet
\lc\ and apply EPD on the \oot\ portion of the follow-up \lcs. We
scale the formal photometric errors on each \lc\ so that $\chi^{2} /
{\rm d.o.f.} = 1$ for the \oot\ portion of the \lc. We take the
magnitudes and radii of the stars from the
\cite{baraffe:1997,baraffe:1998} isochrones, transforming the $BVRI$
magnitudes to the Sloan system using the relations from
\cite{jordi:2006}. To make a fair comparison between the blend models
and the H-p model we also fit the H-p model to the \lcs\ using the
same procedure used to fit the blend-models (see \refsec{globmod} for
a more detailed analysis of the H-p model used for the final parameter
determinations). Fig.~\ref{fig:blend-ht-lcs} compares the best fit
H,S-s, H,s-p and H-p models.

The best fit H,S-s model, consisting of an eclipsing pair with masses
$M_{1} = 0.73\msun$ and $M_{2} = 0.12\msun$ that is diluted by the
star \hatcur{} ($M=\hatcurISOm\,\msun$), has $\chisq = 1445$ with 1333
degrees of freedom. We compare this to the best fit H-p model which
has $\chisq = 1364$ with 1334 degrees of freedom. Because the
photometric noise appears to be temporally correlated
(\reffig{blend-ht-lcs}), formal estimates for the significance of
$\Delta \chisq$ between two models will overestimate the confidence
with which one model can be rejected in favor of another. We therefore
conduct Monte Carlo simulations to estimate the expected distribution
of $\Delta \chisq$ values under the assumption that the H-S,s model is
correct, and accounting for temporal correlations in the noise. To
generate light curves for the Monte Carlo simulations that have
similar time-correlated noise as the real light curves we Fourier
transform the residual of each light curve from the best-fit H,S-s
model, randomize the phases, inverse Fourier transform it, and then
add in the H,S-s model. This method forces the simulated light curves
to have the same noise power-spectrum (and hence auto-correlation
function) as the actual light curve residuals. We scale the errors of
each simulated light curve to have $\chisq / {\rm d.o.f.} = 1$ in the
\oot\ portion of the \lc. The Fourier transforms are carried out
assuming uniform time-sampling (a good approximation for the follow-up
light curves); to use the Fast-Fourier Transform algorithm, we
cyclically repeat each light curve so that the total number of points
in the light curve is a power of 2. We then fit the H-p and H,S-s
models to the simulated sets of light curves and record $\Delta \chisq
= \chisq_{\rm H,p} - \chisq_{\rm H,S-s}$ for each simulation. From
1000 simulations we find a median value of $\Delta \chisq = 2.41$ with
a standard deviation of 12.4; we find no instances where $\Delta
\chisq < -81$, the minimum value attained is $-45$. We conclude that,
based on the light curves, the H,S-s model can be rejected in favor of
the H-p model at the $\ga 6\sigma$ confidence level.

For the H,s-p scenario, we find that the best-fit model consists of a
star with mass $M_{1} = 0.72\msun$ transited by a planet with $R_{\rm
  p} = 1.35\rjup$ and diluted by the star \hatcur{}
($M=\hatcurISOm\,\msun$). This model has $\chisq = 1390$ with 1333 degrees of
freedom. To determine the significance of $\Delta \chisq = \chisq_{\rm
  H,p} - \chisq_{\rm H,S-p} = -26$ we repeat the Monte Carlo
simulations, this time adopting the best-fit H,s-p model as the
fiducial model. From 1000 simulations we find a median value of
$\Delta \chisq = 14.3$ with a standard deviation of 13.9. There are 4
simulations with $\Delta \chisq \leq -26$, so we conclude that the H-p
model is preferred over the H,s-p model with $99.6\%$ ($\la 3\sigma$)
confidence.

As described in \refsec{globmod} it is possible to correct for
systematic errors in the photometry by simultaneously applying EPD and
TFA to the light curves while fitting a physical model to them. By
using a more sophisticated model of this form we are able to rule out
the H,s-p model with higher confidence, and also rule out the H-p,s
model. We perform the global modeling as described in \refsec{globmod}
incorporating three additional parameters that allow for dilution in
the $g$-, $i$- and $z$-bands. This model effectively encompasses both
the H-p,s and H,s-p models because the only H,s-p models that provide
a reasonable fit to the light curve are models where the
planet-bearing star has a mass that is nearly equal to the mass of the
diluting star \hatcur{}. We allow the dilution factors to vary
independently in the fit. We find that models with no dilution are
strongly preferred, and place 1$\sigma$ upper limits on the light
contribution in each filter from an uneclipsed star (``third light'')
of $l_{3,z} < 3\%$, $l_{3,i} < 3.5\%$ and $l_{3,g} < 5\%$. Any
additional star thus makes a negligible contribution to the total
light of the system. This test thus rules out both the H-p,s and H,s-p
models.

\ifthenelse{\boolean{emulateapj}}{\begin{figure*}[]}{\begin{figure}[]}
\ifthenelse{\boolean{emulateapj}}{}{\epsscale{0.9}}
\plotone{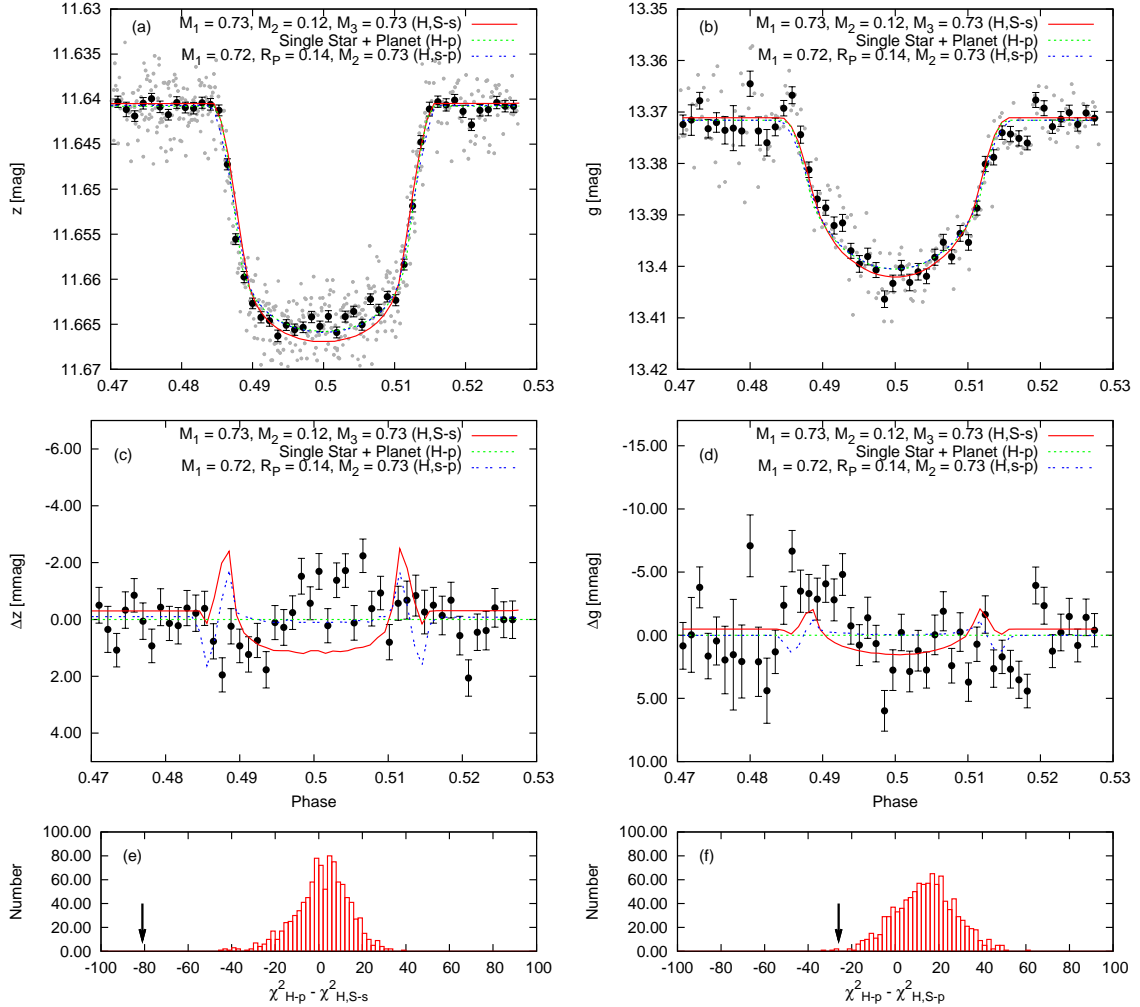}
\caption{
	Hierarchical triple blend model fits to the follow-up \band{z}
        (a) and \band{g} (b) \lcs. EPD filtering was performed on the
        light curves in \oot\ mode (i.e. a transit model was not
        simultaneously fit to the light curves). We compare the
        hierarchical triple models to a model consisting of a single
        star orbited by a planet. The gray-scale points are the
        un-binned data while the dark points are the binned data. Due
        to time correlations in the noise, the scatter in the binned
        data is higher than would be expected for white noise (the
        error bars show the expected errors if the noise were
        white). To save space we only show the $z$- and \band{g} \lcs,
        although the follow-up Sloan \band{i} and HATNet \band{I}
        \lcs\ were also included in the fit. Panels (c) and (d) show
        the residuals about the best-fit H-p model for the follow-up
        $z$- and \band{g} \lcs, respectively. Note that the vertical
        axis for the residual \lcs\ is in mmag. Panels (e) and (f)
        show the distribution of $\chi^{2}_{\rm H-p} - \chi^{2}_{\rm
          H,S-s}$ and $\chi^{2}_{\rm H-p} - \chi^{2}_{\rm H,s-p}$
        respectively from 1000 Monte Carlo simulations of light curves
        with correlated noise. The arrows mark the observed
        values. While the best fit H,S-s and H,s-p models are
        difficult to distinguish by eye from the best fit H-p model,
        they are statistically rejected in favor of the H-p model at
        $\sim 6\sigma$ and $\sim 3\sigma$ confidence respectively.
\label{fig:blend-ht-lcs}}
\ifthenelse{\boolean{emulateapj}}{}{\epsscale{1.0}}
\ifthenelse{\boolean{emulateapj}}{\end{figure*}}{\end{figure}}

\subsubsection{Bisector Variations Induced by Sky Contamination}\label{sec:scf}

As shown in sections~\ref{sec:hitrip} and~\ref{sec:bgeb}, blend
scenarios involving an eclipsing binary star system are inconsistent
with other observations of the system. We therefore look for an
explanation of the apparent BS-RV correlation shown in \refsec{bisec}
that does not invoke a blend. One possibility is that it is due to
varying contamination from the sky spectrum. Because \hatcur{} is
relatively faint, the flux from the sky is non-negligible compared to
the flux from the source. Scattered light from the moon illuminating
the sky near \hatcur{} has a solar-like spectrum which yields a peak
near ${\rm RV}=0\,\kms$ in the cross-correlation profile of the
observed spectrum. The degree to which this second peak contaminates
the peak from the star varies with the sky brightness and the radial
velocity difference between the moon and the star. Because the
observer-centric velocity of \hatcur{} is always less than the
velocity of the moon, an increase in the sky brightness or a decrease
in the velocity difference will lead to a positive BS variation for
our adopted BS sign convention. 

While the sky brightness is not directly measureable from the
available data, in order to quantify this effect we may introduce a
sky contamination factor (SCF) given by
\begin{equation}
{\rm SCF} = \frac{I}{\Delta V^{2} + \left( \frac{1}{2} \Gamma \right)^{2}}
\label{eqn:scf}
\end{equation}
where $I$ is the ratio of the flux in the spectrum due to the moon to
the flux due to the star, $\Delta V$ is the observer-centric radial
velocity difference between the moon and the star, $\Gamma =
15.28\,\kms$ is the width of the Lorentzian function that best fits
the mean cross-correlation profile, and the form for the denominator
is chosen because the cross-correlation profile is well fit by a
Lorentzian function. We estimate $I$ via the relation
\begin{equation}
I = \frac{T_{0}/t_{0}}{T/t}10^{-0.4(B_{m} - B_{S})}
\label{eqn:moonbrightness}
\end{equation}
where $T$ is the total flux received in the region of the spectrum
used to compute the BS, $t$ is the exposure time, $T_{0}$ and $t_{0}$
are the values for the spectrum with the highest count-rate (these are
used to account for changes in the flux received from the star due to
variations in the seeing or transparency), $B_{S}$ is the \band{B}
magnitude of the star (we take $B_{S} = 13.84$ assuming $(B-V) \sim
1.0$ for a dwarf star with $T_{\rm eff} = 4650$K), and $B_{m}$ is the
effective magnitude of the sky due to the moon at the position of the
star (in an area of $\sim 2.5$ square arcseconds). To estimate $B_{m}$
we use the model for the sky brightness due to moonlight given by
\citet{krisciunas:1991}, extending it to the $B$ band by taking $(B -
V) = 0.91$ for the moon \citep[e.g.][]{schaefer:1998} and $k = 0.19$
for the extinction coefficient \citep[this is a typical value at the
  summit of Mauna Kea for the Johnson \band{B} which is roughly the
  region of the spectrum used to calculate the
  BS,][]{krisciunas:1987}. The values for $B_{m}$ range from $18.15$
to $19.86$. When the moon is below the horizon we take $B_{m} =
99.99$.

\begin{figure}[!ht]
\ifthenelse{\boolean{emulateapj}}{}{\epsscale{0.5}}
\plotone{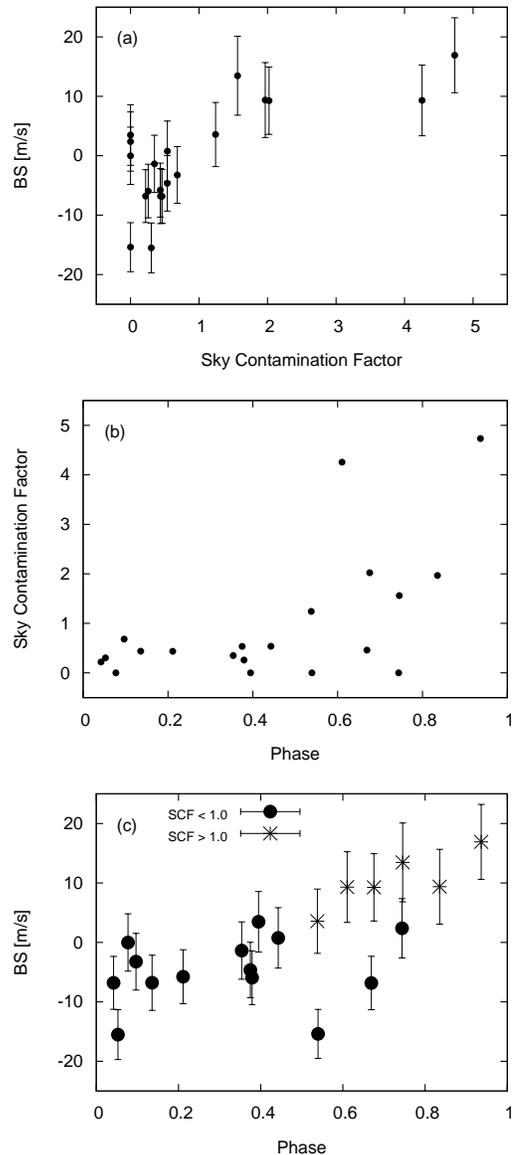}
\caption{
	(a) BS vs. SCF (eq.~\ref{eqn:scf}). The two variables appear
  to be positively correlated. (b) SCF vs. orbital phase. Spectra
  taken between phases 0.5 and 1 had by chance higher sky
  contamination on average than the spectra taken between phases 0 and
  0.5. As a result the BS are correlated with the orbital phase. (c)
  BS vs. orbital phase shown separately for points with high and low
  SCF. When points with high SCF are removed, there is no longer an
  apparent correlation between BS and orbital phase.
\label{fig:scf}}
\ifthenelse{\boolean{emulateapj}}{}{\epsscale{1.0}}
\end{figure}

Fig.~\ref{fig:scf} compares the SCF to the BS values and to the
orbital phase. Note that we normalize the SCF to have a mean value of
1.0. There is a positive correlation between the SCF and the BS. By
chance, spectra taken between orbital phases 0.5 and 1 had higher sky
contamination on average than those taken between orbital phases 0 and
0.5. When points with ${\rm SCF} > 1.0$ are removed, the correlation
between the remaining BS and RV values is no longer significant (the
correlation significance is 37\%). We conclude that this a plausible
explanation for the apparent BS-RV correlation.

\begin{figure}[!ht]
\plotone{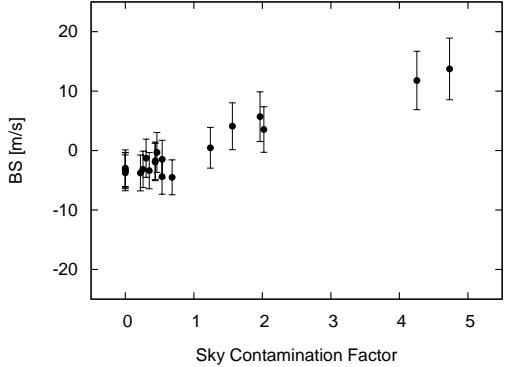}
\caption{
	BS vs. SCF (eq.~\ref{eqn:scf}) for simulated spectra with sky
        contamination. The correlation is similar to that seen for the
        real observations (fig.~\ref{fig:scf}a).
\label{fig:scfsim}}
\end{figure}

As a further test on this hypothesis we simulate sky contaminated
spectra and measure the BS values using the same procedure as for the
actual spectra. To simulate a spectrum we take
\begin{equation}
s_{i} = z(t,V_{t} - V_{i}) + I\times z(t_{2},V_{t_{2}} - \gamma_{t_{2}})
\end{equation}
where $t$ is the iodine-free template spectrum of \hatcur, $t_{2}$ is
the iodine-free template spectrum of HAT-P-13 scaled to have the same
total flux though the \band{B} as $t$ \citep[HAT-P-13 has $T_{\rm eff}
  = 5638$~K, and is thus a better approximation to a solar spectrum
  than \hatcur;][]{bakos:2009b}, $V_{t}$ and $V_{t_{2}}$ are the
barycentric velocity corrections for the templates, $V_{i}$ is the
barycentric velocity correction for spectrum $i$, $\gamma_{t_{2}}$ is
the average radial velocity of HAT-P-13, $I$ is given by
eq.~\ref{eqn:moonbrightness}, and $z(x,y)$ is a function that
redshifts the spectrum $x$ by velocity $y$. Figure~\ref{fig:scfsim}
compares the SCF to the BS for the simulated spectra. The simulations
show a correlation between the SCF and BS that is comparable to that
seen in fig.~\ref{fig:scf}(a). This confirms that sky contamination
may affect the BS values at the level that is observed.

\subsection{Global modeling of the \hatcur{} system}
\label{sec:globmod}

Our model for the follow-up \lcs\ used analytic formulae based on
\citet{mandel:2002} for the eclipse of a star by a planet, where the
stellar flux is described by quadratic limb-darkening. The limb
darkening coefficients were derived from the SME results
(\refsec{stelparam}), using the tables provided by \citet{claret:2004}
for $z$-, $i$-, and $g$-bands. The transit shape was parametrized by the
normalized planetary radius $p\equiv \rpl/\rstar$, the square of the
impact parameter $b^2$, and the reciprocal of the half duration of the
transit $\zrstar$. We chose these parameters because of their simple
geometric meanings and the fact that these show negligible correlations
\citep[see][]{bakos:2009}. Our model for the HATNet data was the simplified
``P1P3'' version of the \citet{mandel:2002} analytic functions, for the
reasons described in \citet{bakos:2009}. 
Following the formalism presented by \citet{pal:2009}, the RV curve
was parametrized by an eccentric Keplerian orbit with semi-amplitude
$K$, and Lagrangian orbital elements
$(k,h)=e\times(\cos\omega,\sin\omega)$.

We assumed that there is a strict periodicity in the individual
transit times. In practice, we assigned the transit number $N_{tr} =
0$ to the first high quality follow-up \lc\ gathered on 2007 March 27.
The adjusted parameters in the fit were the first transit center
observed by HATNet, $T_{c,-132}$, and the last transit center observed
by the \flwof\ telescope, $T_{c,+212}$, covering all of our
measurements with the HATNet telescopes, and the \flwof\ telescope. We
prefer using $T_{c,-132}$ and $T_{c,+212}$ as adjusted parameters
rather than the period and epoch for the reasons discussed by
\citet{bakos:2007c} and \citet{pal:2008a}. The transit center times
for the intermediate transits were interpolated using these two epochs
and the $N_{tr}$ transit number of the actual event. The model for the
RV data contains the ephemeris information through the $T_{c,-132}$
and $T_{c,+212}$ variables \citep{pal:2009}. Altogether, the 11
parameters describing the physical model were $T_{c,-132}$,
$T_{c,+212}$, $\rpl/\rstar$, $b^2$, $\zrstar$, $K$, $k = e\cos\omega$,
$h = e\sin\omega$, and three additional ones related to the
instrumental configuration. These are the instrumental blend factor
$B_{\rm inst}$ of HATNet which accounts for possible dilution of the
transit in the HATNet light curve, the HATNet out-of-transit
magnitude, $M_{\rm 0,HATNet}$, and the relative RV zero-point
$\gamma_{rel}$.

We extended our physical model with an instrumental model that
describes the systematic variations of the data. This was done in a
similar fashion to the analysis presented in \citet{bakos:2009}.
Basically, the HATNet photometry has been already EPD- and
TFA-corrected before the global modeling, so we only considered
systematic corrections to the follow-up \lcs. We chose the ``ELTG''
method, i.e.~EPD was performed in ``local'' mode with EPD coefficients
defined for each night, and TFA was performed in ``global'' mode using
the same set of stars and TFA coefficients for all nights. The
underlying physical model was based on the \citet{mandel:2002}
analytic formulae, as described earlier.  The five EPD parameters were
the hour angle (characterizing a monotonic trend that changes linearly
over time), the square of the hour angle, and the stellar profile
parameters (equivalent to FWHM, elongation, position angle).
The exact functional form of the above parameters contained 6
coefficients, including the auxiliary out-of-transit magnitude of the
individual events. The EPD parameters were independent for all 4
nights, implying 24 additional coefficients in the global fit. For the
global TFA analysis we chose 18 template stars that had good quality
measurements for all nights and on all frames, implying an additional
18 parameters in the fit. We apply EPD to the template star \lcs\ using the
same set of parameters as used for the \hatcur{} \lcs\ before
incoporating them in the analysis. Thus, the total number of fitted
parameters is 11 (physical model) + 24 (local EPD) + 18 (global TFA) =
53, i.e.~much smaller than the number of data-points ($\gtrsim1000$).

The joint fit was performed as described in \citet{bakos:2009}.
We minimized \chisq\ in the parameter space by using a
hybrid algorithm, combining the downhill simplex method \citep[AMOEBA,
  see][]{press:1992} with the classical linear least squares
algorithm.  Uncertainties on the parameters were derived using the
Markov Chain Monte-Carlo method \citep[MCMC, see][]{ford:2006} using
``Hyperplane-CLLS'' chains \citep{bakos:2009}. The \emph{a priori}
distributions of the parameters for these chains were chosen from a
generic Gaussian distribution, with eigenvalues and eigenvectors
derived from the Fisher covariance matrix for the best fit value.  The
Fisher covariance matrix is calculated analytically using the partial
derivatives given by \citet{pal:2008b} and \citet{pal:2009}. Since the
eccentricity of the system appeared as insignificant
($k=\hatcurRVkecc$, $h=\hatcurRVhecc$), we repeated the global fit by
fixing these to 0. The best fit results for the relevant physical
parameters are summarized in
\reftab{planetparam}. \reftab{planetparam} also lists the RV
``jitter'', which is a component of assumed astrophysical noise
intrinsic to the star that we add in quadrature to the RV measurement
uncertainties in order to have $\chi^{2}/{\rm dof} = 1$ from the RV
data for the global fit. In addition, some auxiliary parameters (not
listed in the table) were:
$T_{\mathrm{c},-132}=\hatcurLCTA$~(BJD),
$T_{\mathrm{c},+212}=\hatcurLCTB$~(BJD), $\gamma_{rel}=\hatcurRVgamma$\,\ms
(for the Keck RVs, note that this does {\em not} correspond to the
true center of mass radial velocity of the system, but is only a
relative offset).
The planetary parameters and their uncertainties can be derived by the
direct combination of the \emph{a posteriori} distributions of the \lc,
radial velocity and stellar parameters.  We found that the mass of the
planet is $\mpl=\hatcurPPmlong\,\mjup = \hatcurPPmelong\,\mearth$, the
radius is $\rpl=\hatcurPPrlong\,\rjup = \hatcurPPrelong\,\rearth$ and
its density is $\rho_p=\hatcurPPrho$\,\gcmc. The final planetary
parameters are summarized at the bottom of Table~\ref{tab:planetparam}.

\begin{deluxetable}{lc}
\tablewidth{0pc}
\tablecaption{Orbital and planetary parameters\label{tab:planetparam}}
\tablehead{
	\colhead{~~~~~~~~~~~~~~~Parameter~~~~~~~~~~~~~~~} &
	\colhead{Value}
}
\startdata
\sidehead{\Lc{} parameters}
~~~$P$ (days)             \dotfill    & $\hatcurLCP$              \\
~~~$T_c$ (${\rm BJD}$)    \dotfill    & $\hatcurLCT$              \\
~~~$T_{14}$ (days)
      \tablenotemark{a}   \dotfill    & $\hatcurLCdur$            \\
~~~$T_{12} = T_{34}$ (days)
    \tablenotemark{a}     \dotfill    & $\hatcurLCingdur$         \\
~~~$\arstar$              \dotfill    & $\hatcurPPar$             \\
~~~$\zrstar$              \dotfill    & $\hatcurLCzeta$           \\
~~~$\rpl/\rstar$          \dotfill    & $\hatcurLCrprstar$        \\
~~~$b^2$                  \dotfill    & $\hatcurLCbsq$            \\
~~~$b \equiv a \cos i/\rstar$
                          \dotfill    & $\hatcurLCimp$            \\
~~~$i$ (deg)              \dotfill    & $\hatcurPPi$ \phn         \\

\sidehead{RV parameters}
~~~$K$ (\ms)              \dotfill    & $\hatcurRVK$              \\
~~~$k_{\rm RV}$\tablenotemark{b} 
                          \dotfill    & $\hatcurRVk$              \\
~~~$h_{\rm RV}$\tablenotemark{b}
                          \dotfill    & $\hatcurRVh$              \\
~~~$e$                    \dotfill    & $\hatcurRVeccen$          \\
~~~RV jitter (\ms)           \dotfill    & \hatcurRVjitter           \\

\sidehead{Planetary parameters}
~~~$\mpl$ ($\mjup$)       \dotfill    & $\hatcurPPmlong$          \\
~~~$\rpl$ ($\rjup$)       \dotfill    & $\hatcurPPrlong$          \\
~~~$C(\mpl,\rpl)$
    \tablenotemark{c}     \dotfill    & $\hatcurPPmrcorr$         \\
~~~$\rhopl$ (\gcmc)       \dotfill    & $\hatcurPPrho$            \\
~~~$a$ (AU)               \dotfill    & $\hatcurPParel$           \\
~~~$\log g_p$ (cgs)       \dotfill    & $\hatcurPPlogg$           \\
~~~$T_{\rm eq}$ (K)       \dotfill    & $\hatcurPPteff$           \\
~~~$\Theta$\tablenotemark{d}               \dotfill    & $\hatcurPPtheta$          \\
~~~$\langle F \rangle$ (\trillionergscmsq) \tablenotemark{e}
                          \dotfill    & $\hatcurPPfluxavgnoexp$        \\
\enddata
\tablenotetext{a}{%
	\ensuremath{T_{14}}: total transit duration, time
	between first and last contact; 
	\ensuremath{T_{12}=T_{34}}:
	ingress/egress time, time between first and second, or third and fourth
	contact.}
\tablenotetext{b}{
	Fixed to 0.
}
\tablenotetext{c}{
	Correlation coefficient between the planetary mass \mpl\ and radius
	\rpl.
}
\tablenotetext{d}{
	The Safronov number is given by $\Theta = \frac{1}{2}(V_{\rm esc}/V_{\rm orb})^2 = (a/\rpl)(\mpl / \mstar )$ \citep[see][]{hansen:2007}.
}
\tablenotetext{e}{
	Incoming flux per unit surface area.
}
\end{deluxetable}


\section{Discussion}
\label{sec:discussion}

\ifthenelse{\boolean{emulateapj}}{\begin{figure*}[]}{\begin{figure}[]}
\plotone{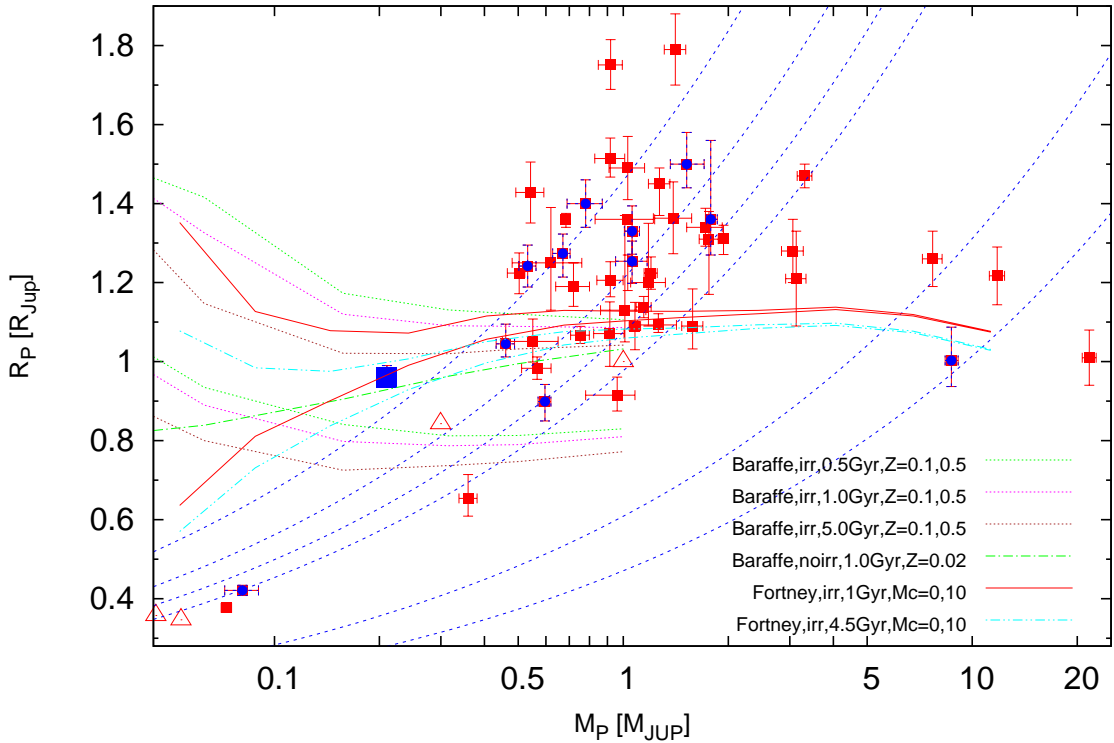}
\caption{ 
  Mass--radius diagram of TEPs (filled squares; blue for those found
  by HATNet and red for those found by other surveys) and solar system
  planets (triangles). \hatcurb\ is shown as a large filled square on
  the left.  Overlaid are the \citet{baraffe:2008} irradiated
  planetary isochrones for ages of 0.5, 1.0 and 5.0~Gyr and metal
  fractions of $Z = 0.1$ and $Z = 0.5$, the \citet{baraffe:2008}
  non-irradiated, 1.0~Gyr, solar metallicity isochrone, and the
  \citet{fortney:2007} 1.0~Gyr and 4.5~Gyr isochrones for planets with
  core masses of $M_{C} = 0$ and $M_{C} = 10\mearth$ interpolated to
  the solar-equivalent semi-major axis of \hatcurb. We also show the
  isodensity lines for 0.4, 0.7, 1.0, 1.33, 5.5 and 11.9\,\gcmc\ (dashed
  lines). \hatcurb\ appears to be well-modeled as a $1.0-4.5$~Gyr,
  mildly irradiated planet with a core mass of $M_{C} \la
  10\mearth$. \hatcurb\ is the lowest-mass H/He dominated gas giant
  planet found to date.
\vspace{0.6cm}
}
\label{fig:exomr}
\ifthenelse{\boolean{emulateapj}}{\end{figure*}}{\end{figure}}

Comparing \hatcurb{} to the theoretical models of \cite{baraffe:2008},
we find that the mass and radius of the planet are consistent with the
1.0~Gyr, $Z = Z_{\odot} = 0.02$ non-irradiated model, or with a
1.0-5.0~Gyr, $0.10 < Z < 0.50$ irradiated model (\reffig{exomr}). With
an equilibrium temperature of $T_{\rm eq} = \hatcurPPteff$\,K,
\hatcurb{} has an equivalent solar semi-major axis of $a_{\rm equiv} =
0.084$~AU, so the irradiation received by the planet, while not
insignificant, is less than what is used to calculate the irradiated
model ($a_{\rm equiv} = 0.045$~AU). The inferred metal fraction is
expected to be closer to $Z = 0.10$ than $Z = 0.50$ for the
\cite{baraffe:2008} models if the correct irradiation were used.

We have also compared \hatcurb{} to the theoretical models of
\cite{fortney:2007}. In \reffig{exomr} we have interpolated these
models to $a_{\rm equiv} = 0.084$~AU, and find that the mass and radius of
\hatcurb{} are consistent with a 10\mearth\ core, 1~Gyr model, and lie
between the core-less and 10\mearth\ core, 4.5~Gyr models. We conclude,
therefore, that \hatcurb{} is most likely a H/He dominated planet with
a core of perhaps $\la 10\mearth$, and a total metal fraction of $\la
15$\%. This makes \hatcurb{} the least massive H/He dominated gas
giant planet found to date; the previous record holder was Saturn.

It is interesting to compare the properties of \hatcurb{} to those of
Saturn and HD~149026b, the two planets with known radii that have
masses closest to that of \hatcurb{}. Measurements of the mass of
HD~149026b range from $0.36\mjup$ to $0.37\mjup$, while determinations
of its radius range from $0.65\rjup$ to $0.813\rjup$
\citep{sato:2005,dc:2006,torres:2008,winn:2008,nutzman:2009,carter:2009}. The
planet appears to have a significant core, with estimates ranging from
$45\mearth$ to $114\mearth$ \citep[see][and references
  therein]{carter:2009}, implying a high metal fraction of $Z \ga
0.4$. Saturn has a mass of $0.299\mjup$ \citep{standish:1995},
equatorial radius of $0.843\rjup$ \citep{seidelmann:2007}, an
estimated core mass of $9\mearth \la M_{C} \la 22\mearth$, and a total
heavy element fraction of $0.14 \la Z \la 0.29$
\citep{saumon:2004}. Although \hatcurb{} is less massive than both
HD~149026b and Saturn, it has a larger radius than both planets. Note
that \hatcurb{} does not have a detectable eccentricity, so its large
radius may not be due to tidal heating \citep[in the models
  by][however, close-in planets may have tidally inflated radii even
  with eccentricities $\la 0.01$]{jackson:2008}. The large radius in
comparison with Saturn may be due in part to the enhanced irradiation
received by \hatcurb{}, and to \hatcurb{} potentially having a smaller
core mass than Saturn. HD~149026b, on the other hand, receives more
irradiation than \hatcurb{} \citep[$a_{\rm equiv} = 0.025$~AU using
  the parameters from][]{carter:2009}, so the difference in radii
suggests that \hatcurb{} has a substantially smaller core mass and
metal enhancement than HD~149026b.

It is interesting to note that the inferred core mass of the three
planets appears to correlate with the host star metallicity
\citep[\hatcur{} has $\feh = \hatcurSMEizfeh$, the Sun has $\feh = 0$,
  and HD~149026 has $\feh = 0.36 \pm 0.05$ from][]{sato:2005}. This
correlation has been previously noted by \citet{guillot:2006} and by
\citet{burrows:2007}, and is perhaps suggestive evidence for the core
accretion model of planet formation \citep[e.g.][and references
  therein]{alibert:2005}. Further discoveries of TEPs with masses
comparable to or less than that of Saturn are needed to determine
whether or not this correlation holds. Note from \reffig{exomr} that
the radii of planets in this mass regime are more sensitive to the
core mass than are the radii of more massive planets for which a given
core mass is a smaller fraction of the total planet mass.

Finally, one might wonder why other planets like \hatcurb{} have not
been found to date (see \reffig{exomr}). The significant $\sim 2.5\%$
transit depth of \hatcurb{} is well within the range that is easily
detectable for many transit surveys, and the \hatcurRVK{}\,\kms RV
semi-amplitude, while small, is still easily measured with
high-precision RV spectrometers (though more observations may be
needed for a robust confirmation, which may slow the rate at which
these planets are announced). We conclude that of hot gaseous planets
with radii similar to Jupiter, only a small fraction have masses
similar to Saturn such as \hatcurb; the majority have masses similar
to Jupiter. With the discovery of \hatcurb, we estimate that the
fraction is $\sim 2\%$, with considerable uncertainty.



\acknowledgements 

We would like to thank the referee, Peter McCullough, for several
suggestions that improved the quality of this paper, and Scott Gaudi
for a helpful discussion. HATNet operations have been funded by NASA
grants NNG04GN74G, NNX08AF23G and SAO IR\&D grants. Work of
G.\'A.B.~and J.~Johnson were supported by the Postdoctoral Fellowship
of the NSF Astronomy and Astrophysics Program (AST-0702843 and
AST-0702821, respectively). We acknowledge partial support also from
the Kepler Mission under NASA Cooperative Agreement NCC2-1390 (D.W.L.,
PI). G.K.~thanks the Hungarian Scientific Research Foundation (OTKA)
for support through grant K-60750. This research has made use of Keck
telescope time granted through NOAO and NASA. The Digitized Sky
Surveys were produced at the Space Telescope Science Institute under
U.S. Government grant NAG W-2166. The images of these surveys are
based on photographic data obtained using the Oschin Schmidt Telescope
on Palomar Mountain and the UK Schmidt Telescope. The plates were
processed into the present compressed digital form with the permission
of these institutions. This research has made use of the SIMBAD
database, operated at CDS, Strasbourg, France.



\end{document}